\documentstyle[epsf]{aipproc}

\begin{document}
\title{Cluster Abundance and\\ Large Scale Structure}
\author{Jiun-Huei Proty Wu}
\address{Astronomy Department,
  University of California, Berkeley,
  CA 94720-3411, USA}

%\lefthead{LEFT head}
%\righthead{RIGHT head}
\maketitle

\vspace*{-10mm}
\begin{abstract}
We use the presently observed number density of large X-ray clusters
and linear mass power spectra to constrain 
the shape parameter ($\Gamma$), the spectral index ($n$),
the amplitude of matter density perturbations
on the scale of $8 h^{-1}$Mpc ($\sigma_8$),
and the redshift distortion parameter ($\beta$).
The non-spherical-collapse model as an improvement to the Press-Schechter 
formula is used and yields significantly lower $\sigma_8$ and $\beta$.
An analytical formalism for the formation redshift of halos
is also derived.
\end{abstract}

%=========================================

One of the most important constraints on models of structure formation is the 
observed abundance of galaxy clusters. 
Because they are the largest virialized objects in the universe,
their abundance can be simply predicted by and thus used to
constrain the linear perturbation theory.
In the light of the new observations and the improvement in modeling
cluster evolution, we revisit this application \cite{caalss},
which has been extensively explored in the literature. %=====cite? HERE

\begin{figure}
  \centering 
  \leavevmode\epsfxsize=2.8in \epsfbox{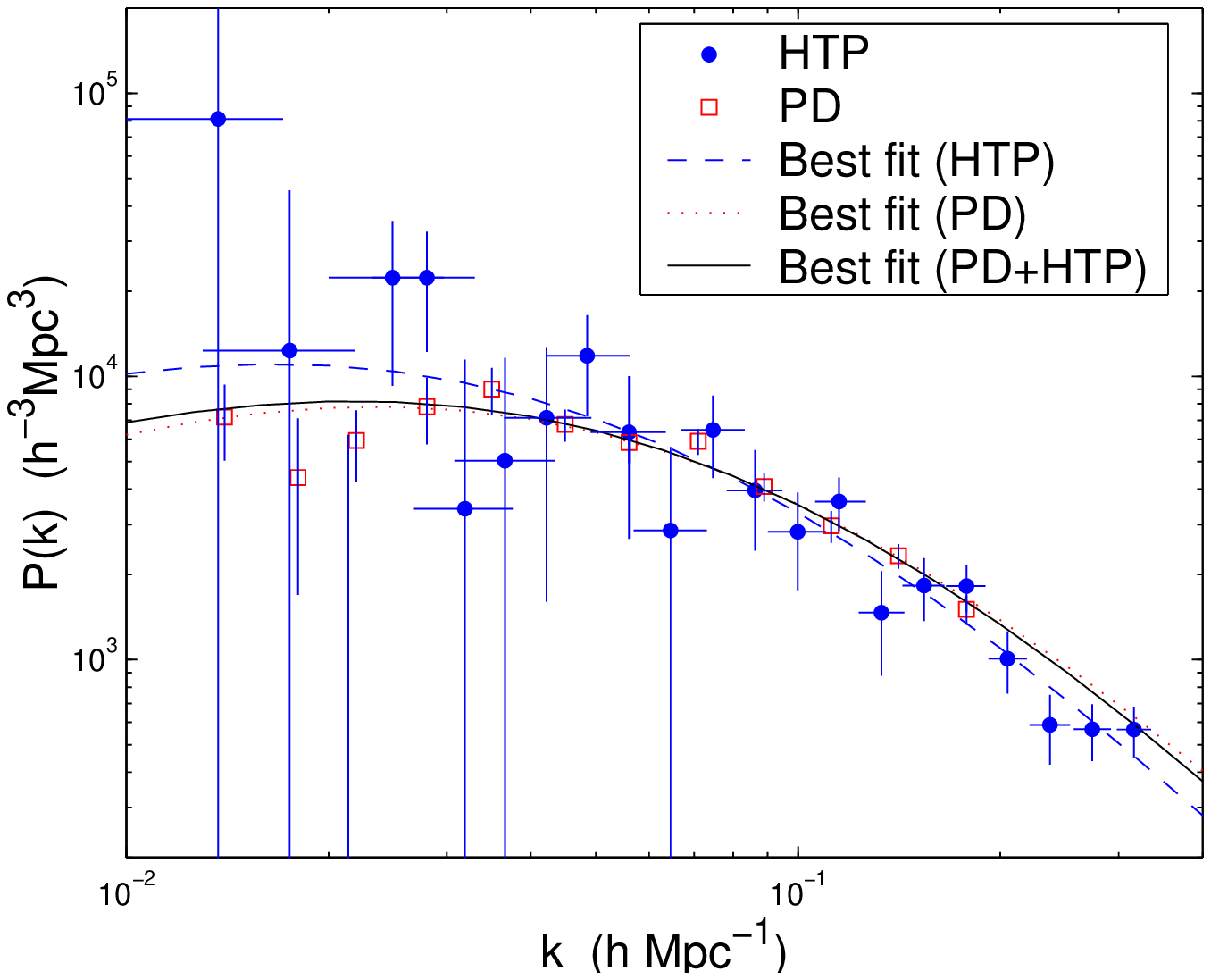}
  \leavevmode\epsfxsize=2.8in \epsfbox{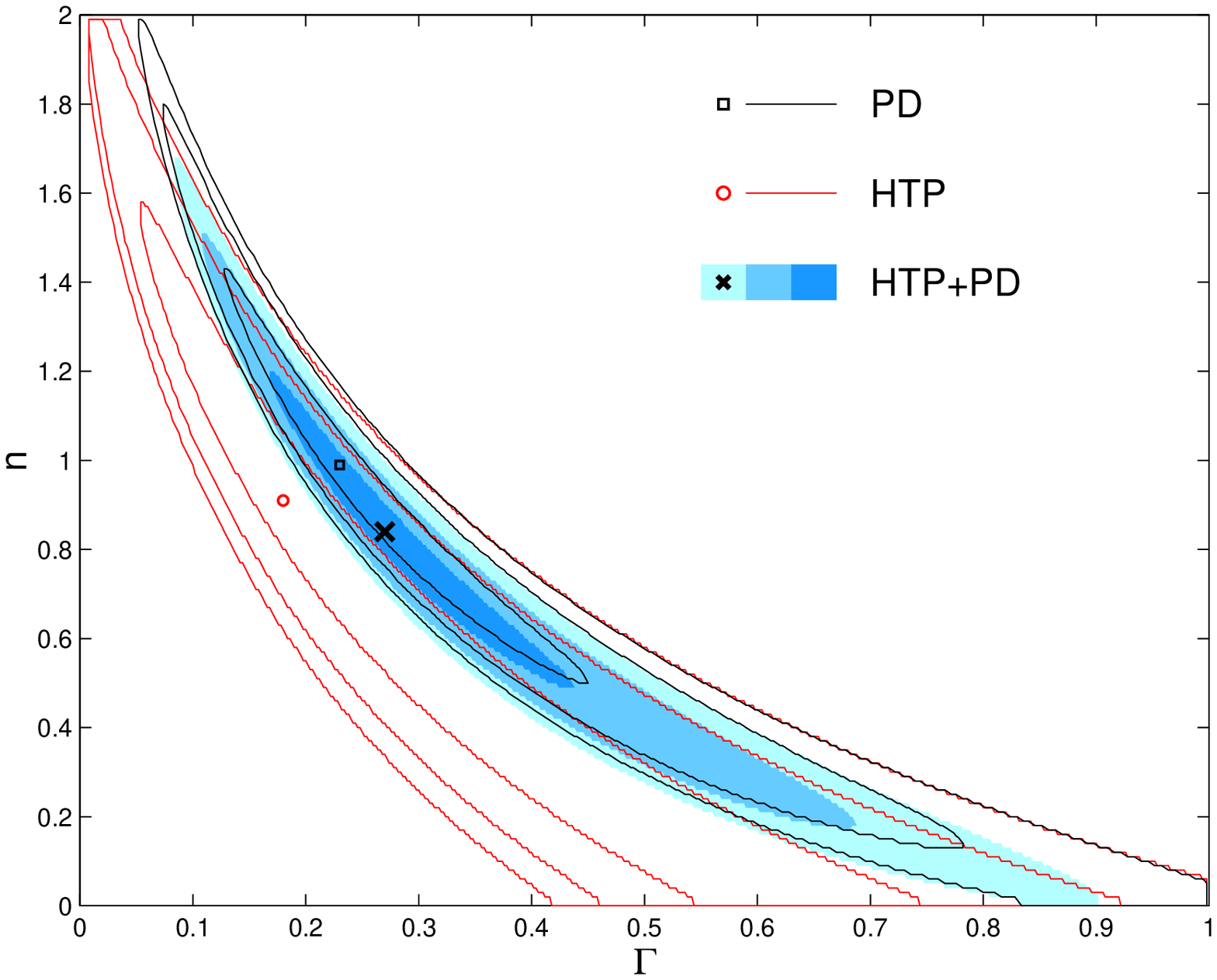}\\
  \caption
  {Matter power spectra of different observations and their best fits (left).
   The 68\%, 95\%, and 99\% likelihood contours (inner out)
   in the $(\Gamma, n)$ parameter space (right).}
  \label{figure1}
\end{figure}

Based on the maximum-likelihood analysis,
we first use the observed linear mass power spectra $P(k)$
by Peacock \& Dodds \cite{PD} (PD, combination of galaxy surveys) and 
by Hamilton, Tegmark, \& Padmanabhan \cite{HTP} 
(HTP, based on PSCz \cite{Saunders2000}; see Figure~\ref{figure1}),
to estimate
the spectral index $n$, the shape parameter $\Gamma$,
and the amplitude of perturbations $\sigma_8$
in the parameterization of the standard model
$P(k)\propto \sigma_8^2 k^n T^2(k/\Gamma)$ \cite{caalss}.
The results are shown in Figure~\ref{figure1} and Table~\ref{table1},
with $\sigma_{8{\rm (I)}}=0.78\pm 0.26$ for IRAS galaxies.
The degeneracy between $\Gamma$ and $n$ is clear,
motivating us to find
the theoretically expected `degenerated' shape parameter 
$\Gamma'=0.247\Gamma\exp(1.4n)=0.220^{+0.036}_{-0.031}$,
which has a much more constrained likelihood.
These results are consistent with the current constraints from CMB 
\cite{maxiboom}.
\begin{table}
\caption{Best fits of different data sets (all errors at 95\% confidence level).}
\label{table1}
\begin{tabular}{lcccc}
 & $n$ & $\Gamma$ & $\Gamma'$ & $\chi^2$/degrees of freedom (conf.\ level)\\
\tableline
    HTP    & $ 0.91^{+1.09}_{-0.91}$ & $ 0.18^{+0.74}_{-0.18}$ 
           & $ 0.160^{+0.085}_{-0.051}$ & 15.4/19 (70\%) \\
\tableline
    PD     & $ 0.99^{+0.81}_{-0.86}$ & $ 0.23^{+0.55}_{-0.16}$
           & $ 0.229^{+0.042}_{-0.033}$ & 6.95/9 (64\%) \\
\tableline
    HTP+PD & $ 0.84^{+0.67}_{-0.67}$ & $ 0.27^{+0.42}_{-0.16}$
           & $ 0.220^{+0.036}_{-0.031}$ & 24.6/30 (74\%) \\
\end{tabular}
\end{table}

\begin{figure}
  \centering 
  \leavevmode\epsfxsize=2.85in \epsfbox{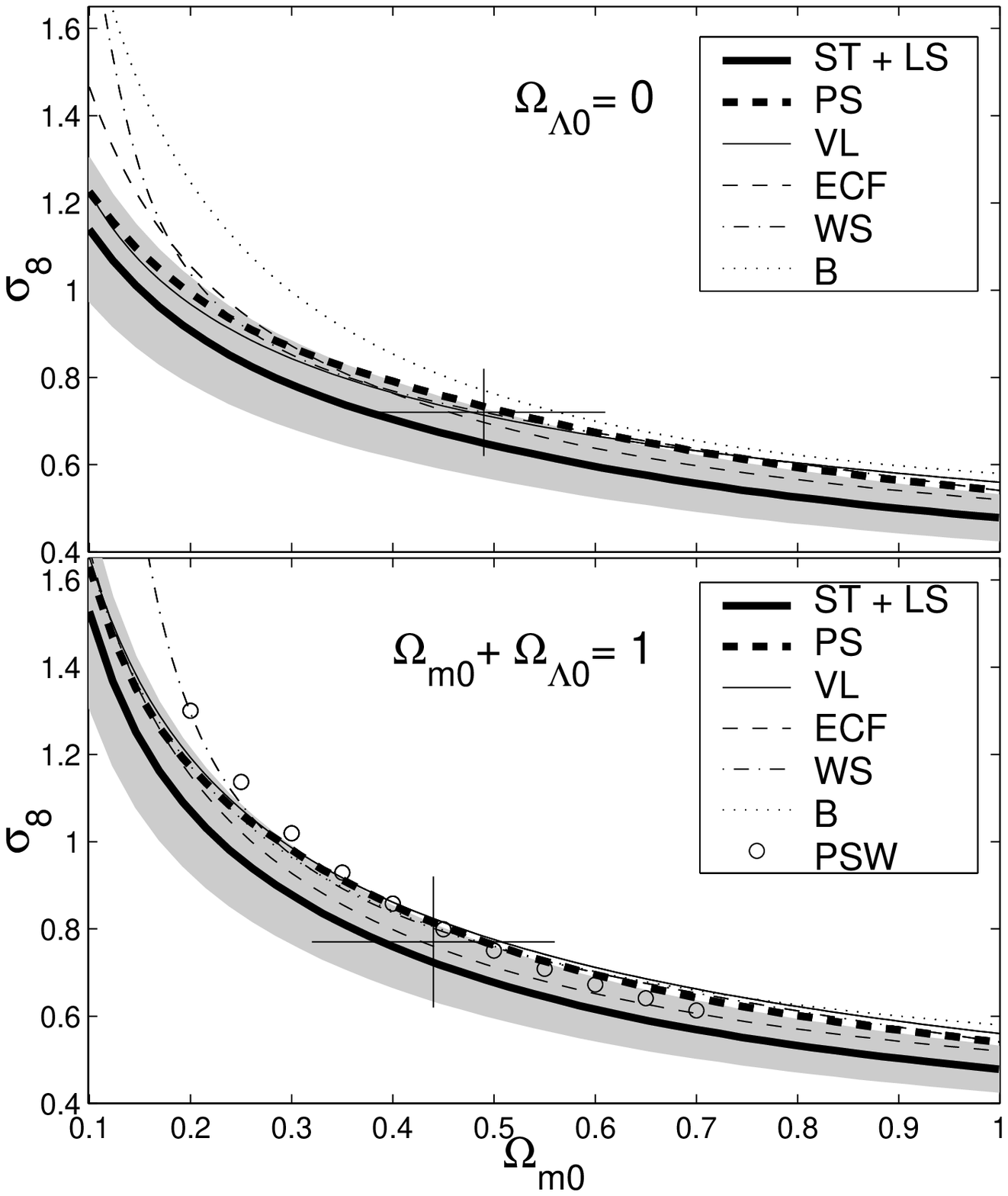}
  \leavevmode\epsfxsize=2.7in \epsfbox{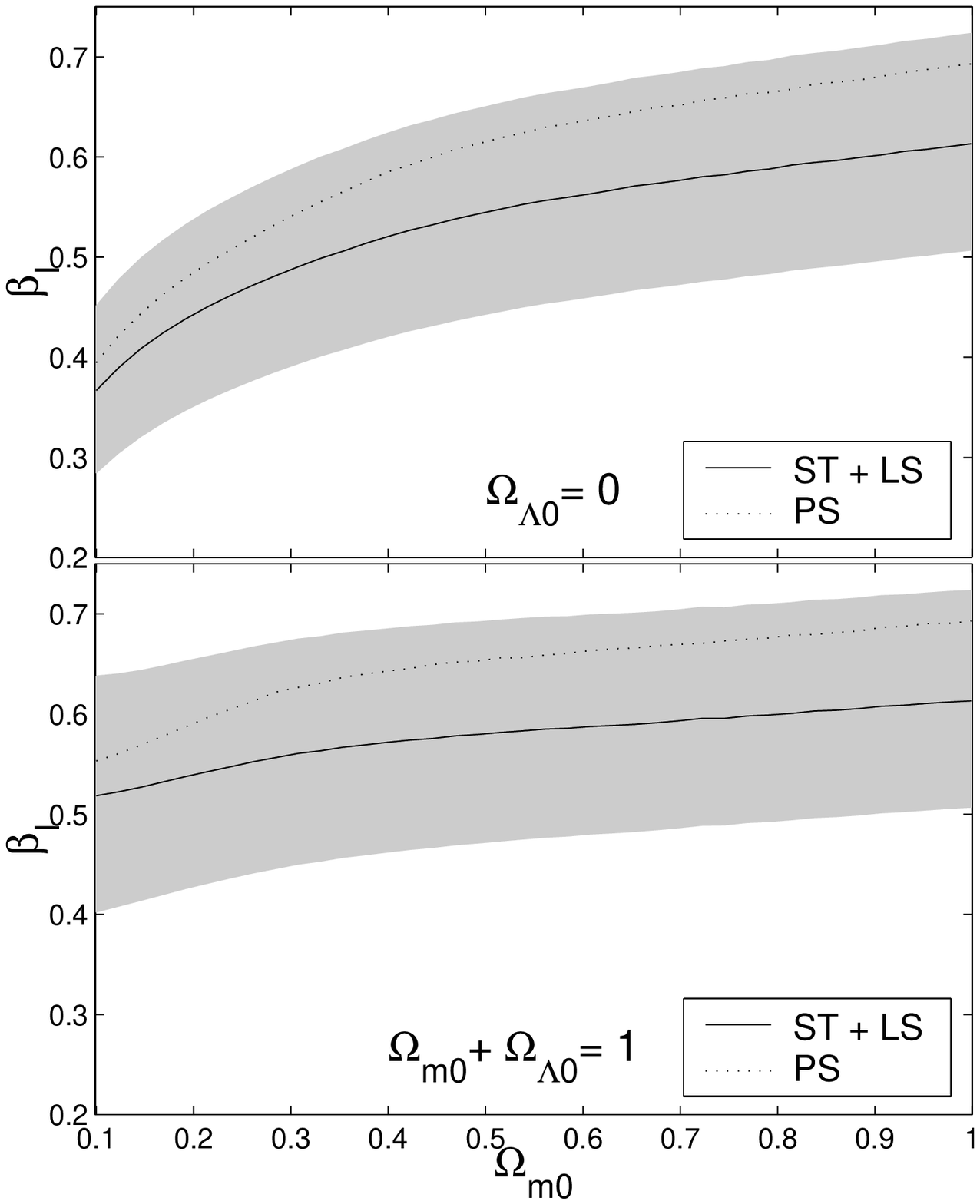}\\
  \caption[]
  {The cluster-abundance-normalized $\sigma_{8}$ and  $\beta_{\rm I}$,
in comparison with results of $\sigma_{8}$ from the literature
(see \cite{s8} for the abbreviations used in the figure legend).
}
  \label{figure4}
\end{figure}

%==============================
Following a similar formalism as in Ref.~\cite{LC},
we then derive
the probability distribution function $p_{z(i)}(z)$
of cluster formation redshift $z$ \cite{caalss}
for different models of mass function $n_i(M)$,
where $i=$PS (Press \& Schechter \cite{PS}),
ST (Sheth \& Tormen \cite{ShethTormen}),
or
LS (Lee \& Shandarin \cite{LeeShandarin1}),
the last two of which incorporate non-spherical collapse .
%(see figure~\ref{figure3}).
%\begin{figure}
%  \centering 
%  \leavevmode\epsfxsize=2.8in \epsfbox{figure2_1.eps}
%  \leavevmode\epsfxsize=2.8in \epsfbox{figure2_2.eps}\\
%  \caption
%  {The mass functions $n_i$ (left)
%   of different models 
%   and their resulting probability distribution $p_{\omega(i)}(\omega)$ (right)
%   of halo formation epochs $\omega(z)$, 
%   where 
%   $p_{\omega(i)}(\omega)d\omega=p_{z(i)}(z)dz$ and 
%   $d\omega/dz>0$.
%   }
%  \label{figure3}
%\end{figure}
With specified $n_i(M)$, $p_{z(i)}(z)$, $\sigma_8$,
and the previously estimated $n$ and $\Gamma$ (or $\Gamma'$),
we can project the present cluster abundance of a given mass $M$
into the space of formation redshift $z$, 
and then use
the virial mass-temperature relation to
associate this abundance with the virial temperature $T$ that
corresponds to the given $M$ and $z$.
An integration over $T$ and $z$ will
give us a prediction of cluster abundance,
which is a function of $\sigma_8$.
A comparison of this with the observation \cite{Henry1} will give
the normalization of $\sigma_8$.
Combined with the $\sigma_{8{\rm (I)}}$ estimated earlier,
it further yields the constraint on 
the redshift distortion parameter
$\beta_{\rm I}\approx \Omega_{\rm m}^{0.6}\sigma_8/ \sigma_{8{\rm (I)}}$,
which quantifies the confusion between the Hubble expansion and 
the local gravitational collapse \cite{Kaiser}.
Our results can be fitted by
$
  \sigma_{8(i)}(\Omega_{\rm m0},\Omega_{\Lambda 0})
    = c_1 \Omega_{\rm m0}^{\alpha}$,
where
$i=$PS, ST, LS or ST+LS, and
$
  \alpha
  \equiv
  \alpha(\Omega_{\rm m0},\Omega_{\Lambda 0}) = 
  -0.3 -0.17 \Omega_{\rm m0}^{c_2} 
  - 0.13 \Omega_{\Lambda 0}
$ (see Figure~\ref{figure4} left),
and 
$
  \beta_{{\rm I} (j)}(\Omega_{\rm m0},\Omega_{\Lambda 0})=
	d_1 \Omega_{\rm m0}^{d_2-0.16(\Omega_{\rm m0}+\Omega_{\Lambda 0})}$,
where
$j=$PS or ST+LS (see Figure~\ref{figure4} right).
The parameter values of these fits are given in table~\ref{table2}.

\begin{table}
\caption{Values in the fits of $\sigma_{8(i)}$ and $\beta_{{\rm I} (j)}$.}
  \label{table2}
\begin{tabular}{c|cccc||c|cc}
    $i$ & PS & ST & LS & ST+LS & $j$ & PS & ST+LS \\
    \hline
    $c_1$ & 0.54 & 0.50  & 0.455 & 0.477& $d_1$ & 0.693 & 0.613\\
    \hline
    $c_2$  & 0.45 & 0.37  & 0.31 & 0.34 & $d_2$ & 0.26 & 0.24 \\
  \end{tabular}
\end{table}

%==========================================

It is clear that 
the $\sigma_8$ and $\beta_{\rm I}$ resulted from
non-spherical-collapse models (ST and LS)
are systematically lower than those based on the PS formalism,
mainly owing to the larger mass function on cluster scales.
A detailed investigation of the uncertainties in our final results
shows that
the main contributor is 
the uncertainty in 
the normalization of the virial mass-temperature relation.
Therefore
further improvement in this normalization
will provide us with more stringent constraint 
on both $\sigma_8$ and $\beta_{\rm I}$.
In addition,
since 
we saw significant corrections in the resulting
$\sigma_8$ and $\beta_{\rm I}$
when switching from the PS formalism
to the more accurate non-spherical-collapse models,
we urge the use of these models in all relevant studies,
especially when we are entering 
the regime of precision cosmology.
We acknowledge the support from 
NSF KDI Grant (9872979) and
NASA LTSA Grant (NAG5-6552).

%==========================================

\end{document}